\documentclass[preprint]{vgtc}               

\graphicspath{{figures/}{pictures/}{images/}{./}} 

\usepackage{times}                     

\usepackage{tabu}                      
\usepackage{booktabs}                  
\usepackage{lipsum}                    
\usepackage{mwe}                       

\usepackage{mathptmx}                  

\onlineid{0}

\vgtccategory{Research}

\vgtcinsertpkg

\usepackage{cuted} 
\usepackage{ccicons} 

\usepackage{orcidlink} 

\usepackage{xcolor} 
\usepackage{tcolorbox} 
\usepackage{fancybox} 
\usepackage{appendix} 
\usepackage{wrapfig} 
\usepackage{graphicx}
\usepackage{ragged2e} 

\newcommand{\eg}{e.\,g.}
\newcommand{\ie}{i.\,e.}
\newcommand{\etal}{et al.\@\xspace}

\newcommand{\done}[1]{\textcolor{black}{#1}}

\usepackage{makecell}

\definecolor{myPurple}{RGB}{134,132,190}
\definecolor{myCyan}{RGB}{142,183,165}
\definecolor{myOrange}{RGB}{242,164,60}
\definecolor{myRed}{RGB}{182,88,88}
\definecolor{myTextPurple}{RGB}{80,79,114}
\definecolor{myTextCyan}{RGB}{82,106,96}
\definecolor{myTextOrange}{RGB}{140,95,35}
\definecolor{myTextRed}{RGB}{156,75,75}

\newcommand{\ShadeTaskTarget}[1]{%
  \begin{Sbox}
    \mbox{#1}
  \end{Sbox}
  \setlength{\fboxrule}{0pt} 
  \setlength{\fboxsep}{1pt} 
  \fcolorbox{black}{myPurple!40}{\TheSbox}
}
\newcommand{\ShadeUser}[1]{%
  \begin{Sbox}
    \mbox{#1}
  \end{Sbox}
  \setlength{\fboxrule}{0pt} 
  \setlength{\fboxsep}{1pt} 
  \fcolorbox{black}{myCyan!40}{\TheSbox}
}
\newcommand{\ShadeWidget}[1]{%
  \begin{Sbox}
    \mbox{#1}
  \end{Sbox}
  \setlength{\fboxrule}{0pt} 
  \setlength{\fboxsep}{1pt} 
  \fcolorbox{black}{myOrange!40}{\TheSbox}
}
\newcommand{\ShadeRelationship}[1]{%
  \begin{Sbox}
    \mbox{#1}
  \end{Sbox}
  \setlength{\fboxrule}{0pt} 
  \setlength{\fboxsep}{1pt} 
  \fcolorbox{black}{myRed!40}{\TheSbox}
}
\newcommand{\ShadeSquareExample}[1]{%
  \hspace*{-10pt}
  \begin{Sbox}
    \mbox{\hspace*{1pt}#1\hspace*{-1pt}}
  \end{Sbox}
  \hspace*{1pt}
  \setlength{\fboxrule}{0pt} 
  \setlength{\fboxsep}{1pt} 
  \fcolorbox{black}{myPurple!20}{\TheSbox}
}

\newcommand{\revise}[1]{\textcolor{black}{#1}}

\def\barheight{0.6em}
\def\barwidth{1.5em}

\newcommand{\myBarChart}[2]{
    \kern-0.2em\begin{tikzpicture}[baseline=0ex]
        \fill[white] (0,0) rectangle (\barwidth,\barheight);
        \fill[black!60] (0,0) rectangle ({\barwidth*(#1/#2)},\barheight); 
        \draw[black] (0,0) rectangle (\barwidth,\barheight);
    \end{tikzpicture}
}

\preprinttext{To appear in 2025 IEEE Visualization and Visual Analytics (VIS).}

\ieeedoi{xx.xxxx/TVCG.201x.xxxxxxx}

\newcommand{\papertitle}{Designing Visualization Widgets for Tangible Data Exploration:\\A Systematic Review}

\title{\papertitle}

\author{%
  Haonan Yao \thanks{e-mail: haonanyao0622@gmail.com \orcidlink{0009-0003-1735-6281} }  \hspace{20 pt}
  Lingyun Yu \thanks{e-mail: lingyun.yu@xjtlu.edu.cn \orcidlink{0000-0002-3152-2587} } \hspace{20 pt}
  Lijie Yao \thanks{Corresponding author, 
  e-mail: yaolijie0219@gmail.com \orcidlink{0000-0002-4208-5140} }\\
  \parbox{5in}{\scriptsize \centering Department of Computing, School of Advanced Technology, Xi'an Jiaotong-Liverpool University}
}

\teaser{
  \centering
  \includegraphics[width=\linewidth]{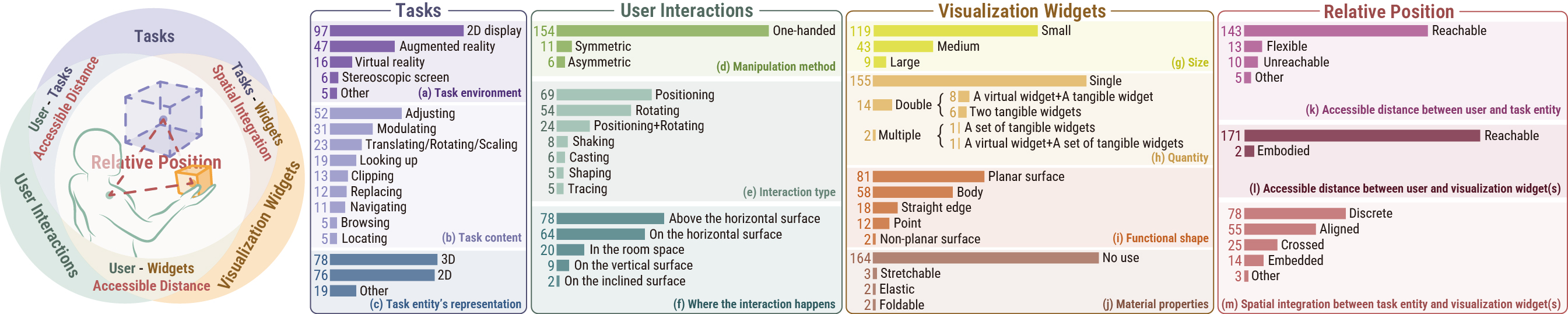}
  \caption{An illustration of the dimensions and their sub-categories in our systematic review. A detailed version is in Appendix~\ref{appx:review}.}
  \label{fig:Teaser}
}

\abstract{
We present a systematic review on tasks, interactions, and visualization widgets \revise{(refer to tangible entities that are used to accomplish data exploration tasks through specific interactions)} in the context of tangible data exploration. 
Tangible widgets have been shown to reduce cognitive load, enable more natural interactions, and support the completion of complex data exploration tasks. Yet, the field lacks a structured understanding of how task types, interaction methods, and widget designs are coordinated, limiting the ability to identify recurring design patterns and opportunities for innovation.
To address this gap, we conduct a systematic review to analyze existing work and characterize the current design of data exploration tasks, interactions, and tangible visualization widgets. We next reflect based on our findings and propose a research agenda to inform the development of a future widget design toolkit for tangible data exploration.
Our systematic review and supplemental materials are available at \revise{\href{https://physicalviswidget.github.io/}{physicalviswidget.github.io} and} \href{https://osf.io/vjw5e/}{osf.io/vjw5e/}.
} 


\begin{document}



\maketitle

\section{Introduction}

Exploring data requires users to complete specific tasks through interactions with visual representations~\cite{Dimara:2019:WhatIsInteraction}. 
A wide range of interaction paradigms have been proposed, including tactile and pen-based input~\cite{Jo:2017:BlendingWIMPandTactile, Yu:2012:TactileInteraction}, mid-air gestures~\cite{Kostic:2024:GestureInteraction, Cavallo:2019:GestureInteraction}, and hybrid approaches combining voice or multi-modal input~\cite{Tsang:2002:VocalInteraction, Hall:2022:VocalInteraction}. 
Among these, tangible interaction—where users manipulate physical objects to control virtual counterparts—has shown particular promise by leveraging users’ natural motor skills and spatial reasoning~\cite{Besançon:2021:3DVisTasks}. 
Specifically, in data exploration, visualization widgets, such as sliders, cutting planes, and data probes, have been widely used as interactive tools to control how data is displayed and interpreted~\cite{Willett:2007:ScentedWidgets, Leithinger:2013:CuttingPlane, Steinman:2000:ParticleSeeding}. 

Despite the wide interest in tangible interactions, we still lack a structured understanding of how task types, interaction techniques, and visualization widgets are coordinated in data exploration. In the absence of such an overview, it remains challenging to identify recurring design patterns or extract fruitful insights for future widget toolkit development. While prior work has explored individual visualization widgets or specific interaction methods, no studies have synthesized this knowledge to examine how tangible widgets, user interactions, and visualization tasks are interlinked.

To fill in this gap, we conduct a systematic review of 71 papers, resulting in 171 instances of data exploration tasks involving tangible visualization widgets and corresponding user interactions. We analyze these instances across four dimensions:\linebreak[4]\raisebox{-0.25\baselineskip}{\includegraphics[height=\baselineskip]{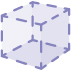}}\hspace*{-3pt}\ShadeTaskTarget{\textbf{Tasks}}\nolinebreak, ~\raisebox{-0.25\baselineskip}{\includegraphics[height=\baselineskip]{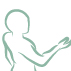}}\hspace*{-3pt}\ShadeUser{\textbf{User Interactions}}\nolinebreak,~\raisebox{-0.25\baselineskip}{\includegraphics[height=\baselineskip]{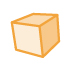}}\hspace*{-3pt}\ShadeWidget{\textbf{Visualization Widgets}}\nolinebreak, and their~\raisebox{-0.25\baselineskip}{\includegraphics[height=\baselineskip]{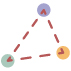}}\hspace*{-3pt}\ShadeRelationship{\textbf{Relative Position}}\nolinebreak.  We further reflect on the findings and propose a research agenda that moves from current practices toward the creation of a tangible widget toolkit.
Our research agenda highlights the need to define reusable functional shapes of widgets, leverage material properties for interaction, and establish deeper coordination between tasks, visualization widgets, and user interactions in data exploration.

\section{Related Work}
As our review lies in the tasks and interactions in data exploration, focusing on the tangible visualization widgets, we first introduce the related work on visualization widgets, followed by visualization tasks and interaction techniques.  

\subsection{Visualization Widgets}
Visualization widgets support interaction by being manipulated by users to control how data is visualized \done{~\cite{Besançon:2021:3DVisTasks}}. It can be virtual or tangible entities. For example, adjusting the transparency of bar charts through a virtual slider on a 2D screen~\cite{Willett:2007:ScentedWidgets}, \done{viewing the interior of a volume rendering by using a virtual cutting plane on a 2D screen~\cite{Beyer:2007:VirtualCuttingPlane, Klein:2012:VirtualCuttingPlane, Rieder:2008:VirtualCuttingPlane}, cutting a volume rendering to intercept slices with a paper above the tabletop~\cite{Spindler:2009:PaperLens}, or replacing the representation of spatio-temporal data via cubes assembly on the table~\cite{He:2024:DataCubesInHand}.}
However, how the physical properties of a tangible widget can be used to benefit the data exploration tasks has not been systematically investigated. A most relevant work to ours is from Gong~\etal~\cite {Gong:2023:AffordanceBased}, highlighting that features like affordances could be leveraged in tangible surface interactions.

\subsection{Visualization Tasks}
In the visualization community and broader \revise{human-computer interaction (HCI)}, data exploration tasks have been classified according to different aspects, such as data characteristics (\eg, \done{graph data tasks~\cite{Lee:2006:GraphTask, Kerracher:2015:GraphTask}} in \revise{information visualization (InfoVis)} and \done{spatial 3D data tasks~\cite{Besançon:2021:3DVisTasks, Keefe:2013:SciVisDataTasks}} scientific vis), interaction tasks (\ie, \done{tasks of controlling virtual entities through actions~\cite{Brudy:2019:InteractionTask, Hertel:2021:InteractionTask, Scoditti:2011:InteractionTask}}), and high-level tasks (\ie, \done{abstract tasks oriented by user intent}~\cite{Besançon:2021:3DVisTasks, Brehmer:2013:AbstractVisTasks}) and low-level tasks (\ie, \done{concrete tasks for accomplishing high-level data visualization and exploration tasks}~\cite{Keefe:2013:SciVisDataTasks}).
In our work, instead of proposing a new classification, we categorize the tasks according to the environment in which they occur, their content, and the representation of the task entity to help clarify relationships between tasks, interactions, and widgets.

\subsection{Interaction Techniques}
Interaction is a goal-oriented activity~\cite{Dimara:2019:WhatIsInteraction} and widely exists in data exploration to help users explore and get insights from data. 
\revise{Yi \etal~\cite{Yi:2007:InfoVisDataTasks} proposed general categories of interaction techniques widely used in Infovis. Specifically}
, Gomez~\etal~\cite{Gomez:2010:FiducialBased} used a cube that can be held by one hand to manipulate data rendering; Ens~\etal~\cite{Ens:2020:Uplift} designed building models that could be placed on a tabletop interface to visualize energy usage inside the building. More recently, He~\etal~\cite {He:2024:DataCubesInHand} proposed a set of flexible cubes that allow users to browse datasets by holding, comparing them by overlaying, or merging them through side-by-side placement. 
Previous work illustrates the diverse roles of tangible widgets in supporting interaction, which our review systematically analyzes to uncover broader design patterns and inform future widget-based data exploration.

\section{Approaches}
In this section, we first introduce how we conducted our systematic review, and next, how we analyzed the results. 

\subsection{Literature Selection and Screening}
\noindent\textbf{Selection:}
We selected papers from major visualization-related and HCI venues, including VIS, EuroVis, CHI, UIST, ISMAR, VR, VRST, DIS, ISS, TEI, SUI, 3DUI, TVCG, CGF, and CG\&A.
Our initial search used the keywords ``tangible \& interaction'' and ``widget'', 
\revise{yielding $611$ papers. We }
later expanded to include ``3D interaction'' after observing that many relevant works in 3D environments employed visualization widgets for data exploration without explicitly referencing \revise{the two keywords above, and} also included ``physical interaction'' to ensure broader coverage of related papers\revise{, resulting in $466$ papers}. 
\revise{We in total selected $1,077$ publications,} spanned from $2010$ to $2024$.

\vspace{2pt}
\noindent\textbf{Screening:}
We excluded papers that lacked a tangible widget (e.g., gesture-only interactions, and  \revise{virtual visualization widgets}), did not involve interaction with a task entity (e.g., annotation without task engagement), or used objects as proxies without further manipulation (e.g., simple pointing). 
We then examined the remaining papers to confirm they clearly described their data exploration tasks, the interactions to complete the tasks, and the (tangible) visualization widgets used. \revise{Papers lacking this information were excluded.}
In total, we \revise{retained $71$ papers} that met our criteria, \revise{including $50$ from keywords on ``tangible \& interaction'', $3$ from ``widget'', and $18$ with ``3D interaction''.}


\subsection{Review and Categorization Process}
We analyzed the $71$ retained papers through a three-phase process:

\vspace{2pt}
\noindent\textbf{Dimension Initialization:}
We defined an initial set of three dimensions aligned with our research scope: \textit{Tasks}, \textit{User Interactions}, and \textit{Visualization Widgets}. Based on our observations during the literature screening phase, that much relevant work had been conducted in an immersive or 3D environment, we thus included \textit{Relative Position} as a fourth dimension to capture the spatial relationships among the user, widget, and task entity. 
Two authors collaboratively reviewed a sample of approximately $15$ papers to define sub-categories for each dimension. One author then applied the resulting categorization scheme to all $71$ papers.

\vspace{2pt}
\noindent\textbf{Dimension Iteration:}
During categorization, if new potential categories emerged, the coding author brought them to the group discussion. All authors jointly evaluated whether to incorporate the new category and clarified its scope. We refer to this as an iterative process. After each iteration, the categorizing author reapplied the updated scheme to the full paper set. Iteration concluded once no additional subcategories emerged.

\vspace{2pt}
\noindent\textbf{Categorization Checking:} 
To ensure clarity and consistency, a second author---independent from the initial coder---reviewed the full categorization to identify ambiguous cases. The two authors resolved disagreements through discussion, and any unresolved cases were brought to the full author group for final decision. The final version of our categorization can be found on OSF (\href{https://osf.io/vjw5e/}{osf.io/vjw5e/}).

As one paper may involve multiple tasks (and thus multiple interactions and widgets), we collected and categorized $171$ instances of tasks from $71$ papers in the end.

\section{Results}
In this section, we describe the results of our systematic review, per dimension and its sub-categories (illustration in \autoref{fig:Teaser}). We use in-line visualization as a visual indicator for major counts.

\vspace{1ex}
\setlength\intextsep{0pt}  
\setlength\columnsep{0pt}  
\begin{wrapfigure}{l}{2\baselineskip}  
    \includegraphics[height=2em]{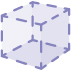}
\end{wrapfigure}
\noindent 
\ShadeTaskTarget{\textbf{Tasks}}describe in which environment the data exploration tasks have been conducted (\textit{Task environment}), what the task content is (\textit{Task content}), and in which representation the task entity has been perceived by users (\textit{Task entity's representation}).

\vspace{1ex}
\noindent\textit{Task environment:} Most data exploration tasks were conducted on 2D displays~\myBarChart{97}{171}, followed by immersive environments such as AR and VR~\myBarChart{63}{171}. In contrast, stereoscopic screens were rarely used; one possible reason is that there have been no dedicated design guidelines on how to show data effectively on such screens.

\vspace{1ex}
\noindent\textit{Task content:} Our collected instances included general tasks such as adjusting~\myBarChart{52}{171}, modulating~\myBarChart{31}{171}, translating/rotating/scaling~\myBarChart{23}{171}, looking up, clipping, replacing, navigating, browsing, and locating. We did not identify other tasks beyond the known types.

\vspace{1ex}
\noindent\textit{Task entity's representation:} 3D~\myBarChart{78}{171} and 2D~\myBarChart{76}{171} are both common representations of task entities; each accounts for almost half of the whole. \revise{Other formats, such as audio, system, artifact, are rare.} We provided a full version in the Appendix~\ref{appx:representation}, demonstrating the detailed representations included in these three types.

\vspace{1ex}
\setlength\intextsep{0pt}  
\setlength\columnsep{0pt}  
\begin{wrapfigure}{l}{2\baselineskip}  
    \includegraphics[height=2em]{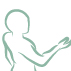}
\end{wrapfigure}
\noindent 
\ShadeUser{\textbf{User Interactions}}describe how users operate tangible entities to complete a data exploration task (\textit{Manipulation method}), what interactions users perform (\textit{Interaction type}), and where interactions take place (\textit{Where the interaction happens}).  

\vspace{1ex}
\noindent\textit{Manipulation method:} In current practices, one-handed manipulation is mainstream~\myBarChart{154}{171}. For two-handed manipulation, symmetric interaction appears approximately twice as often as asymmetric. 

\vspace{1ex}
\noindent\textit{Interaction type:} We categorized the actions users performed into six types:~\raisebox{-0.25\baselineskip}{\includegraphics[height=\baselineskip]{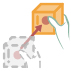}} positioning~\myBarChart{69}{171},~\raisebox{-0.25\baselineskip}{\includegraphics[height=\baselineskip]{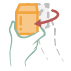}} rotating~\myBarChart{54}{171},~\raisebox{-0.25\baselineskip}{\includegraphics[height=\baselineskip]{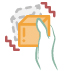}} shaking,~\raisebox{-0.25\baselineskip}{\includegraphics[height=\baselineskip]{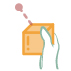}} casting (pointing an object toward a target),~\raisebox{-0.25\baselineskip}{\includegraphics[height=\baselineskip]{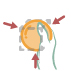}} shaping (modifying an object’s geometry or structure), and~\raisebox{-0.25\baselineskip}{\includegraphics[height=\baselineskip]{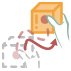}} tracing (drawing a motion trajectory). 
In addition to individual interactions, combination also appeared, and there are quite a few, such as~\raisebox{-0.25\baselineskip}{\includegraphics[height=\baselineskip]{Figures/icon/Interaction_type/positioning.jpg}} positioning $+$ \raisebox{-0.25\baselineskip}{\includegraphics[height=\baselineskip]{Figures/icon/Interaction_type/rotating.jpg}}rotating~\myBarChart{24}{171}.

\vspace{1ex}
\noindent\textit{Where the interaction happens:} Interactions typically took place above horizontal surfaces~\myBarChart{78}{173} (\eg, a tabletop), or on the horizontal surfaces~\myBarChart{64}{173}. 
Some interactions happened in a room space that does not necessarily require a tabletop \myBarChart{20}{173}. A few interactions occurred on vertical and inclined surfaces.  

\vspace{1ex}
\setlength\intextsep{0pt}  
\setlength\columnsep{0pt}  
\begin{wrapfigure}{l}{2\baselineskip}  
    \includegraphics[height=2em]{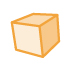}
\end{wrapfigure}
\noindent 
\ShadeWidget{\textbf{Visualization Widgets}}refer to tangible entities that are used to accomplish data exploration tasks through specific interactions. 
\revise{Widget examples include a VR pointer for selecting points via ray-casting, a physical dial/button for adjusting visualization scale, and a pen for positioning.} Here we discuss how big a widget is (\textit{Size}), how many widgets are used (\textit{Quantity}), the shape in which the widget acts (\textit{Functional shape}), and the material properties of the widget (\textit{Material properties}).

\vspace{1ex}
\noindent\textit{Size:} Small widgets—those that fit within a user’s palm—were the most commonly used~\myBarChart{119}{171}. Medium widgets, which exceed palm size but can still be held in one hand, appeared less frequently~\myBarChart{43}{171}. Large widgets, which cannot be grasped by hand, were rarely used. 

\vspace{1ex}
\noindent\textit{Quantity:} Most instances involved a single tangible widget to perform a task~\myBarChart{155}{171}, with occasional use of more than one. In some cases (mainly the clipping task), combined setups were used—either integrating virtual and tangible widgets (e.g.,~\done{a virtual cutting plane controlled by a slider is used to slice a volume rendering accurately, and multiple tops of columns are used for forming a non-planar surface to slice a volume rendering flexibly.})

\vspace{1ex}
\noindent\textit{Functional shape:} 
While widgets may exhibit diverse physical forms, the shape that makes sense might be the same (\eg,~\done{one corner of a tablet is used as point input to select a virtual object, and a distal end of a slender body is used as point input to locate a position in space.}). The most common functional shape was a planar surface~\myBarChart{81}{171}, followed by a body-shaped form~\myBarChart{58}{171}, a straight edge, a point, and a non-planar surface.

\vspace{1ex}
\noindent\textit{Material properties:} Material Properties influenced by the materials from which the widgets are constructed, including~\raisebox{-0.25\baselineskip}{\includegraphics[height=\baselineskip]{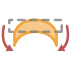}} bendable,~\raisebox{-0.25\baselineskip}{\includegraphics[height=\baselineskip]{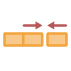}} assembleable,~\raisebox{-0.25\baselineskip}{\includegraphics[height=\baselineskip]{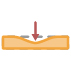}} elastic,~\raisebox{-0.25\baselineskip}{\includegraphics[height=\baselineskip]{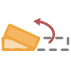}} foldable,~\raisebox{-0.25\baselineskip}{\includegraphics[height=\baselineskip]{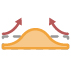}} malleable,~\raisebox{-0.25\baselineskip}{\includegraphics[height=\baselineskip]{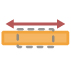}} stretchable, and~\raisebox{-0.25\baselineskip}{\includegraphics[height=\baselineskip]{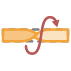}} twistable. 
In our review, most visualization widgets did not leverage material properties~\myBarChart{164}{171}. Only a small number incorporated material-dependent behaviors, such as stretchability, elasticity, or foldability.

\vspace{1ex}
\setlength\intextsep{0pt}  
\setlength\columnsep{0pt}  
\begin{wrapfigure}{l}{2\baselineskip}  
    \includegraphics[height=2em]{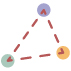}
\end{wrapfigure}
\noindent 
\ShadeRelationship{\textbf{Relative Position}} indicates the  \textit{Accessible distance between user and task entity}, the \textit{Accessible distance between user and visualization widget(s)},  and the \textit{Spatial integration between task entity and visualization widget(s)}.

\vspace{1ex}
\noindent\textit{Accessible distance between user and task entity:} Reachable distance—where the task entity lies within arm’s reach—was the most common configuration in our review~\myBarChart{143}{171}. Less frequent were unreachable distances, flexible distances (with a variable user–entity relationship), and cases where distance could not be determined, such as when the task entity was formless.

\vspace{1ex}
\noindent\textit{Accessible distance between user and visualization widget(s):} Similar to above, reachable distance—where the widget is within the user’s arm’s length—was the most common configuration~\myBarChart{171}{173}. In a few cases, widgets were used in an embodied manner (e.g.,~\done{a bracelet is worn on the wrist to manipulate a 3D virtual object directly}).

\vspace{1ex}
\noindent\textit{Spatial integration between task entity and visualization widget(s):} 
The most frequent configuration was
\raisebox{-0.25\baselineskip}{\includegraphics[height=\baselineskip]{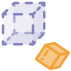}} discrete~\myBarChart{78}{175}, meaning the widget is completely separated from the task entity; and aligned~\myBarChart{55}{175}, indicating side-by-side placement.
Less common were
\raisebox{-0.25\baselineskip}{\includegraphics[height=\baselineskip]{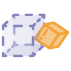}} crossed, where the widget partially overlaps with the task entity, and
\raisebox{-0.25\baselineskip}{\includegraphics[height=\baselineskip]{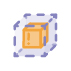}} embedded, where one is fully contained within the other.
“Other” denotes cases where the relative position was unmeasurable, such as when the task entity was formless.

\vspace{1ex}
\textbf{In summary}, we investigate tangible data exploration across four dimensions. We found that most tasks are performed in 2D or immersive environments, with positioning and rotating as the most common interaction types. Interactions are typically one-handed and occur above or on horizontal surfaces. Widgets are generally small, used individually, and without using material properties---most often shaped as planar or body-like forms. Interaction tends to take place within arm’s reach, with task entities and widgets commonly arranged in discrete or aligned spatial integration.

\section{Reflections and A Research Agenda}
Based on the findings of our systematic review, here we discuss our reflections on the current design and outline research opportunities in tangible data exploration.

\subsection{Reflections}
\label{sec:reflections}
Our reflections go from a low level to a high level. As our goal is to inform tangible visualization widget design, we first reflect on the widget design. We next reflect on how the widgets function in interactions to complete concrete tasks. 

\vspace{2ex}
\noindent\textbf{Widget Design:} 

\begin{figure}[t]
    \centering
    \includegraphics[width=\linewidth]{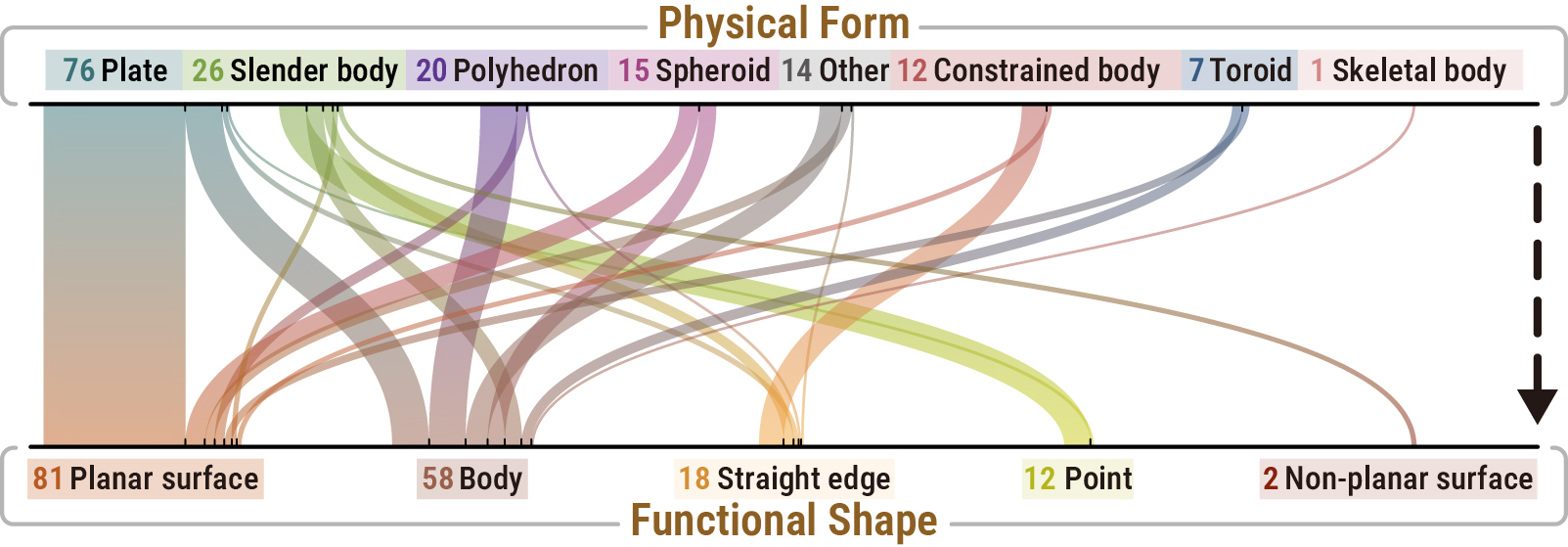}
    \caption{A mapping relationship between the widgets' physical form and functional shape. The definitions of the physical form and functional shape can be found in Appendix~\ref{appx:shape}.}
    \label{fig:ShapeAnalysis}
\end{figure}

\vspace{1ex}
\noindent\textit{The role of shape:} In our systematic review, we introduce \textit{functional shape} of a widget, referring to the geometry relevant to its interactive role. Here, we examine how a widget’s physical form relates to its functional shape (\autoref{fig:ShapeAnalysis}). 
This relationship is not one-to-one: multiple physical forms, such as polyhedrons or slender bodies, may serve the same functional role, and a single form may support multiple functions. These patterns suggest that designers often repurpose different geometries to achieve similar interaction goals. This raises the question: could a general-purpose widget or a toolkit covering key functional shapes streamline the design of tangible tools for data exploration?

\vspace{1ex}
\noindent\textit{Utilization of material properties:} A key distinction between tangible and non-tangible widgets lies in their material-based properties, such as bendability, stretchability, and twistability. Yet, our review reveals that these properties are rarely leveraged in data exploration contexts. For example, although paper is commonly used for straight-line cuts, practitioners often rely on external right-angled objects rather than folding the paper itself to achieve a precise corner. This underuse may stem from a lack of understanding about how such physical properties can be effectively exploited—--and to what extent they support specific data exploration tasks.

\begin{figure*}[t]
    \centering
    \includegraphics[width=\textwidth]{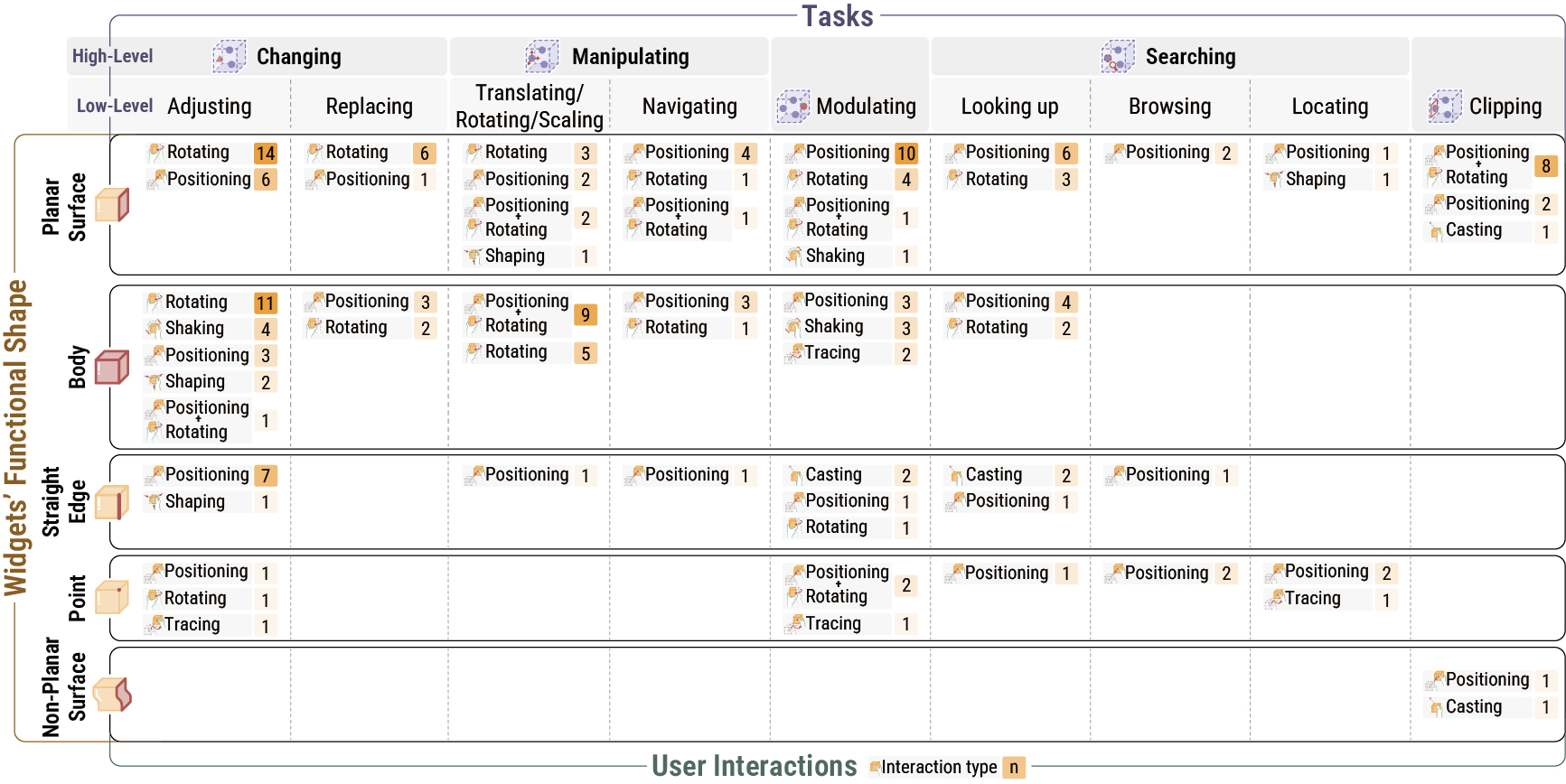}
    \caption{A mapping relationship among the functional shapes of widgets, the types of interactions, and the tasks they support.}
    \label{fig:DesignSpace}
\end{figure*}

\vspace{2ex}
\noindent\textbf{Task-Interaction-Widget as A Whole:} 

\vspace{1ex}
\noindent\textit{Tasks summarization:} As we introduced in related work, visualization tasks have been well classified according to different aspects, such as low-level and high-level tasks. 
In contrast to creating a new task classification, \revise{we summarize the low-level tasks categorization presented in our systematic review under a framework with the high-level tasks that they belong to,} referring to Munzner's~\cite{Munzner:2014:VisualizationBook} and Besançon's~\cite{Besançon:2021:3DVisTasks} classifications. 
They are: \textit{changing}, including replacing a visual representation or adjusting a visual parameter; \textit{manipulating}, including translating/rotating/scaling a visualization or navigating; \textit{searching}, including looking up, browsing, and locating; and \revise{modulating} and \revise{clipping}.

\revise{Our task summarization focuses specifically on tangible visualization widgets, and as such, differs in scope and granularity from prior task classifications in visualization~\cite{Yi:2007:InfoVisDataTasks, Besançon:2021:3DVisTasks, Munzner:2014:VisualizationBook}. We see value in providing this tailored summary, as a lack of existing guidance on which tasks, under which context, have been practically supported by tangible visualization widgets. We draw our summarization to elicit physicalized visualization widget designs in mixed reality.}

\vspace{1ex}
\noindent\textit{Mapping among tasks, interactions, and widgets:} As discussed above, while the physical forms of widgets vary, their functional shapes often converge. 
%
%
To inform the design of tangible visualization widgets, a key is to understand how these functional shapes contribute to interaction and task completion. To this end, we map the functional shapes of widgets to the types of interactions they afford and the tasks they support (\autoref{fig:DesignSpace}).
Specifically, on the widget hand, planar surfaces are highly versatile---used across nearly all tasks with various interactions. Bodies are frequently supporting tasks that involve transformation, like translating, rotating, or scaling, and interactions such as shaking and tracing. Straight edges typically appear in changing tasks and positioning interactions. In contrast, point and non-planar surfaces are rarely used, mapped to fewer task types and interaction styles.  
Regarding interactions, positioning appears across nearly all widget functional shapes and tasks, while other interaction types (\eg, casting, tracing) are tightly coupled to specific widgets' functional shapes. 
In terms of tasks, several tasks lack diverse widget shapes as well as interaction types, which indicate opportunities for future design and research---especially in expanding tangible visualization widgets and interaction techniques.

\subsection{A Research Agenda }
Building on our reflections, we propose a research agenda with a target to inform the design of tangible visualization widgets.

\vspace{1ex} 
\noindent\textbf{Designing a Set of Functional Components:}
Our review reveals that different physical forms often converge toward a limited set of functional shapes, suggesting the potential for general-purpose widget designs. A set of general components that embody core functional shapes, such as planar surfaces and bodies, could save time on widget designing when conducting data exploration tasks and enable designers to prototype tangible interactions quickly. Further work is required towards this end to investigate what components should be included and what their physical forms look like.

\vspace{1ex}
\noindent\textbf{Leveraging Material-Based Physical Properties:} 
Despite tangible widgets inherently possessing diverse material properties, these characteristics are rarely leveraged in current designs. This underuse points to a lack of design guidelines and empirical understanding of how material properties such as blendability, stretchability, and twistability can contribute to interaction and task performance. Further investigations on how materials' physical properties may be purposefully used in data exploration workflows and what new interaction techniques they can enable are required.

\vspace{1ex} 
\noindent\textbf{Toward a Widget Toolkit:}
Our review shows that while widgets' functional shapes like planar surfaces and bodies are widely applicable across tasks and interaction types, others, such as points and non-planar surfaces, remain underutilized. Interactions like positioning are common, whereas others like casting or tracing are shape-dependent and narrowly applied. Although our mappings reveal broad trends among widgets, interactions, and tasks, our current understanding remains fragmented. How tasks, interactions, and widget forms coordinate as an ecological system is still missing, which is an important step toward advancing a widget toolkit for data exploration\revise{, especially for mixed reality contexts where tangible interactions happen the most frequently}. Thus, further work, such as ideation workshops and evaluations, is required to articulate how a toolkit can flexibly support interactions in different task categories.

\section{Conclusion}
We conducted a systematic review of tangible data exploration, focusing on the visualization widgets used, the interactions they afford, and the tasks they support. Building on our review, we reflected on how the physical form and material properties of widgets influence interaction and task performance, and examined the interdependencies among widgets, tasks, and interactions. We conclude with a research agenda outlining three directions for future work. Our work informs current practices in widget design and lays the groundwork for developing a physicalized widget toolkit to support data exploration.

\section*{Supplemental Materials}

\revise{An interactive website of our systematic review is accessible at \href{https://physicalviswidget.github.io/}{physicalviswidget.github.io}}. The supplemental materials for our study are available on OSF at \href{https://osf.io/vjw5e/}{osf.io/vjw5e/}. 
In particular, we share our (a) systematic review material, (b) coding co-occurrences derived from our analysis, and (c) definitions of the physical form and functional shape with examples from our review.





\acknowledgments{
Lijie Yao is partially funded by the XJTLU RDF, grant \textnumero\ RDF-24-01-062. Lingyun Yu was partially supported by the National Natural Science Foundation of China, grant \textnumero\ 62272396.}

\bibliographystyle{abbrv-doi}

\bibliography{PhysicalWidget}

\clearpage

\centering
\begin{strip}
    \centering
    \huge \textbf{\papertitle}\\[.25em]
    \large \textbf{Appendix}\\[.75em]
\end{strip}
\noindent\justifying

In this appendix, we offer additional material, including~\ref{appx:review}. a full version of the dimensions summarized from our systematic review,~\ref{appx:representation}. a full version of the detailed task entity's representation dimension, and~\ref{appx:shape}. definitions of the physical form and functional shape with examples from our review.

\appendix

\section{Detailed Version of Dimensions in the Systematic Review}
\label{appx:review}

\begin{figure*}[ht]
    \centering
    \includegraphics[width=0.8\textwidth]{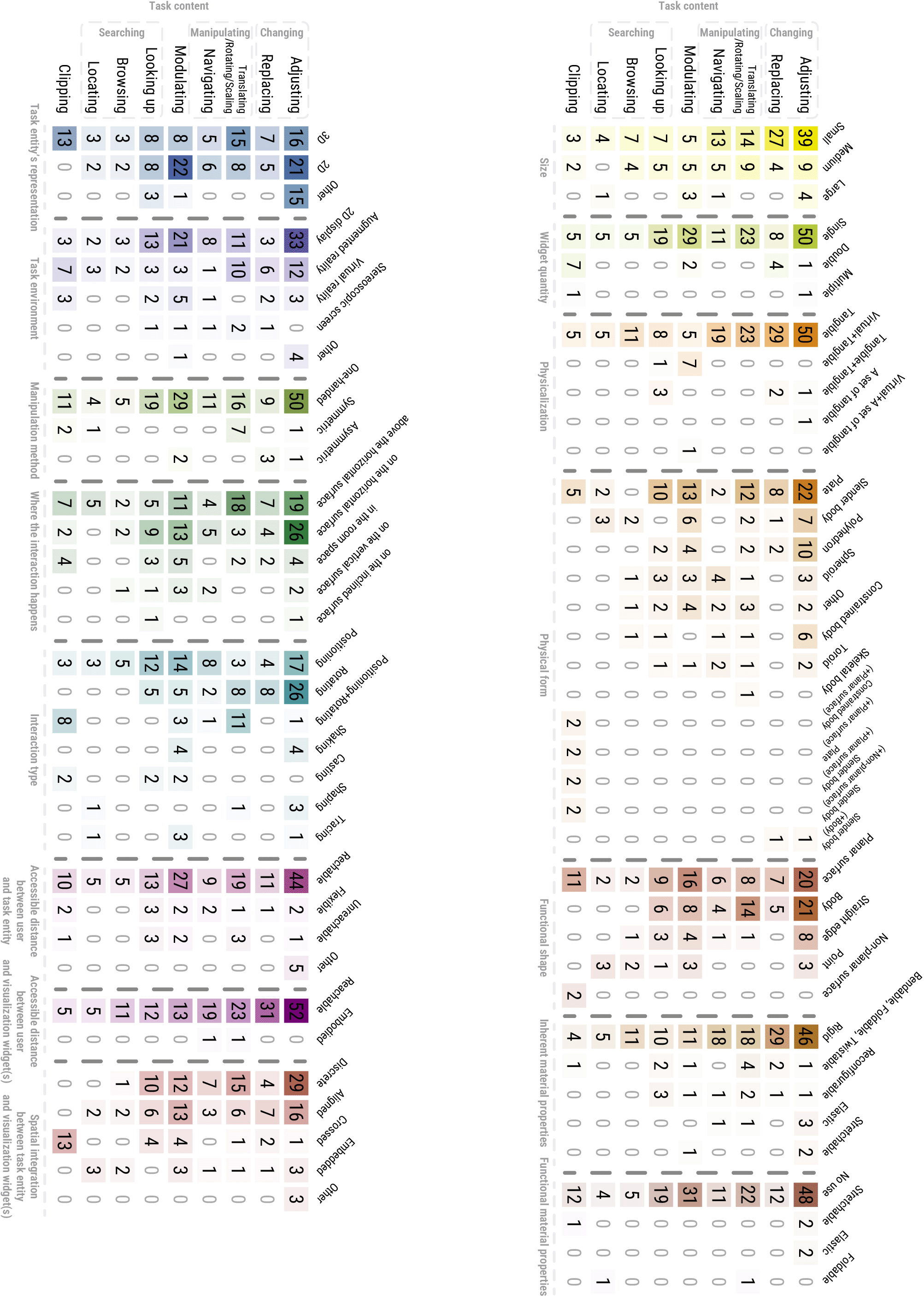}
    \caption{The coding co-occurrences of dimensions summarized from our systematic review, constructed with task content on the y-axis and the other summarized categories on the x-axis. For each item identified by the intersection of the y-axis and x-axis, a color-coded square box \ShadeSquareExample{n} indicates the frequency of its occurrence in the review. The number within the box represents the frequency and the color intensity correlates to the frequency, with darker colors denoting higher frequency counts.}
    \label{fig:Count}
\end{figure*}

Figure~\ref{fig:Count} shows a detailed version of the dimensions summarized from our systematic review, revealing the relationship between task content and each of the other dimensions.

\section{Detailed Version of Task Entity's Representation}
\label{appx:representation}

\begin{figure}[ht]
    \centering
    \includegraphics[width=0.8\columnwidth]{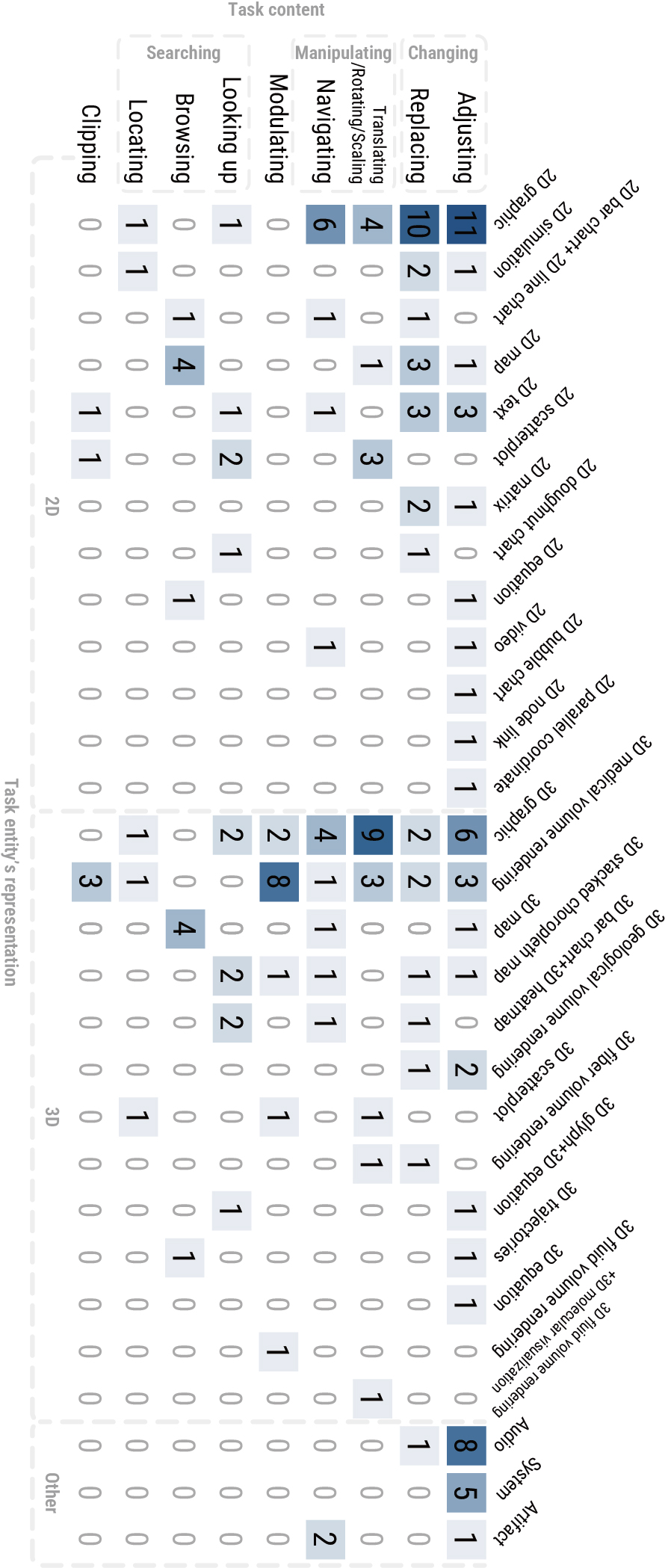}
    \caption{The coding co-occurrences of the tasks entity's representation dimension, constructed with task content on the y-axis and the detailed representations summarized in 2D, 3D, and other on the x-axis. For each item identified by the intersection of the y-axis and x-axis, a color-coded square box \ShadeSquareExample{n} indicates the frequency of its occurrence in the review. The number within the box represents the frequency and the color intensity correlates to the frequency, with darker colors denoting higher frequency counts.}
    \label{fig:DetailedRepresentation}
\end{figure}

Figure~\ref{fig:DetailedRepresentation} show a detailed version of task entity's representation, most task entities referred to visualizations~\myBarChart{154}{171}, other examples included audios, systems, and artifacts that encode data. The non-visualization entities were mainly used for adjusting tasks.
Despite virtual graphics or objects~\myBarChart{59}{171}, 2D visualization involved various representations, but volume rendering predominated in 3D visualization~\myBarChart{21}{171}.

\section{Definitions of the Physical Form and Functional Shape with Examples}
\label{appx:shape}

We listed here the definitions of physical form and functional shape of widgets.

\vspace{2ex}
\noindent\textbf{Physical Form:} We classified the physical form of widgets in terms of shape characteristics and structure:

\vspace{1ex}
\setlength\intextsep{0pt}  
\setlength\columnsep{0pt}  
\begin{wrapfigure}{l}{2\baselineskip}  
    \includegraphics[height=2em]{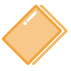}
\end{wrapfigure}
\noindent 
\textit{Plate} is specified by its thickness, which is obviously smaller than its other edges, such as length and width.
This shape extends the features from edges to surface, and it can usually be divided into a polygonal plate (with endpoints and straight edges) and a circular plate (with at least one curved edge). 
Various forms are included, such as a puck, a knob, a fridge sticker, a card for small sizes, and a paper, a board, a tablet for medium sizes.

\vspace{1ex}
\setlength\intextsep{0pt}  
\setlength\columnsep{0pt}  
\begin{wrapfigure}{l}{2\baselineskip}  
    \includegraphics[height=2em]{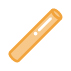}
\end{wrapfigure}
\noindent 
\textit{Slender body} is defined as a shape with small cross-sectional dimensions compared to its length.
This shape has distinctive features due to at least two well-defined endpoints and an edge. Its elongated form provides clear directionality and ease of grasping. 
Examples include a stick that can be pinched between fingers, a stylus matching traditional pen-grip gesture, a baton, or even a controller that is held by curling fingers around its non-planar surface.

\vspace{1ex}
\setlength\intextsep{0pt}  
\setlength\columnsep{0pt}  
\begin{wrapfigure}{l}{2\baselineskip}  
    \includegraphics[height=2em]{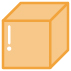}
\end{wrapfigure}
\noindent 
\textit{Polyhedron} is a shape bounded by a finite number of polygons, which has no significant length difference between its edges. 
The polyhedron further extends the features of the shape from surface to body, leading to the ability to proxy an object and ease of manipulation. 
For instance, a pyramid, a cube, a prism, and a cuboctahedron are cases of this shape.

\vspace{1ex}
\setlength\intextsep{0pt}  
\setlength\columnsep{0pt}  
\begin{wrapfigure}{l}{2\baselineskip}  
    \includegraphics[height=2em]{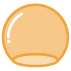}
\end{wrapfigure}
\noindent 
\textit{Spheroid} 
is adopted as the definition of an approximately spherical body, instead of a strict definition of the ellipsoid.
This shape is easy to manipulate, but has no distinct endpoints or edges, surrounded by a continuous non-planar surface. 
Samples of the spheroid include a sphere, a rounded eraser, a door grip, and even a dome with a small bottom.

\vspace{1ex}
\setlength\intextsep{0pt}  
\setlength\columnsep{0pt}  
\begin{wrapfigure}{l}{2\baselineskip}  
    \includegraphics[height=2em]{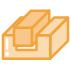}
\end{wrapfigure}
\noindent 
\textit{Constrained body} is an integrated shape formed with a movable part and a constrained part, in which the movable part can move within the physical restriction of the constrained part.
This pair can be physically integrated or separated. Due to the functional close interdependency between the two parts, as discussed by Ullmer~\etal~\cite{Ullmer:2005:ConstrainedBody}. We regard the separable pair still as one constrained body. 
The cases of this shape include, but are not limited to, a stick with a slideable ring and a groove with a slider.

\vspace{1ex}
\setlength\intextsep{0pt}  
\setlength\columnsep{0pt}  
\begin{wrapfigure}{l}{2\baselineskip}  
    \includegraphics[height=2em]{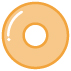}
\end{wrapfigure}
\noindent 
\textit{Toroid} is characterized as a continuous structure with a small overall thickness and a central hole, obtained by rotating a cross-section around the central axis.
The central hole allows the shape to have both internal and external features, and can be further subdivided according to the type of edges and surfaces. 
Examples of this shape include a ring as a lens metaphor, a wearable bracelet, and a torus circling a region of interest.

\vspace{1ex}
\setlength\intextsep{0pt}  
\setlength\columnsep{0pt}  
\begin{wrapfigure}{l}{2\baselineskip}  
    \includegraphics[height=2em]{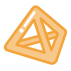}
\end{wrapfigure}
\noindent 
\textit{Skeletal body} is a continuous structure with spatially formed hollows, which can be the outer contour of a form or the spatial structure of the volume's interior.
The holes bring out the spatial internal features and the external features of the entire shape. 
One example is a cubic frame.

\vspace{1ex}
\setlength\intextsep{0pt}  
\setlength\columnsep{0pt}  
\begin{wrapfigure}{l}{2\baselineskip}  
    \includegraphics[height=2em]{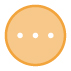}
\end{wrapfigure}
\noindent 
\textit{Other} refers to shapes that go beyond the basic types described above.
These shapes can include a model of a specific object or a shape with complex surfaces obtained through shaping. Their features vary significantly depending on their own diverse structures. 
Given that we aim to abstract the commonality of the physical form of widgets, we do not further classify these shapes that fall outside the defined scope.

\vspace{2ex}
\textbf{Functional Shape:} We deconstructed the elements composing 3D geometric shapes to obtain the functional shapes:

\vspace{1ex}
\setlength\intextsep{0pt}  
\setlength\columnsep{0pt}  
\begin{wrapfigure}{l}{2\baselineskip}  
    \includegraphics[height=2em]{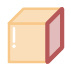}
\end{wrapfigure}
\noindent 
\textit{Planar surface} is a geometric element characterized by its flatness on the shape. 
The flatness describes an overall planarity that does not strictly require constant normal vectors across its extent.

\vspace{1ex}
\setlength\intextsep{0pt}  
\setlength\columnsep{0pt}  
\begin{wrapfigure}{l}{2\baselineskip}  
    \includegraphics[height=2em]{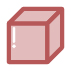}
\end{wrapfigure}
\noindent 
\textit{Body} is defined as the entirety of a three-dimensional geometric shape. 
It is highlighted by the volume that the shape occupies in space.

\vspace{1ex}
\setlength\intextsep{0pt}  
\setlength\columnsep{0pt}  
\begin{wrapfigure}{l}{2\baselineskip}  
    \includegraphics[height=2em]{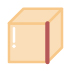}
\end{wrapfigure}
\noindent 
\textit{Straight edge} refers to an element of a geometric shape characterized by linearity. 
The scope of this definition includes the intersection line of planar surfaces and the straight generatrix of a ruled surface.

\vspace{1ex}
\setlength\intextsep{0pt}  
\setlength\columnsep{0pt}  
\begin{wrapfigure}{l}{2\baselineskip}  
    \includegraphics[height=2em]{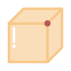}
\end{wrapfigure}
\noindent 
\textit{Point} indicates an element that is a visually prominent portion of a geometric shape. 
In particular, this term encompasses the intersections of edges and the endpoints of protrusions on non-planar surfaces.

\vspace{1ex}
\setlength\intextsep{0pt}  
\setlength\columnsep{0pt}  
\begin{wrapfigure}{l}{2\baselineskip}  
    \includegraphics[height=2em]{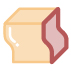}
\end{wrapfigure}
\noindent 
\textit{Non-planar surface} denotes the undulating element on a geometric shape. 
Specifically, undulation refers to exhibiting curvature or irregularity that is not limited to a single convex or concave surface.


\end{document}